\documentclass[letterpaper, 10 pt, journal]{ieeeconf} 
\usepackage[utf8]{inputenc}
\usepackage{amsmath,amssymb,amsfonts}
\usepackage{graphicx}

\title{MIndGrasp: A New Training and Testing Framework for Motor Imagery Based 3-Dimensional Assistive Robotic Control}
\author{Daniel Freer, Guang-Zhong Yang \thanks{D. Freer is with the Hamlyn Centre, South Kensington Campus, Imperial College London, London, SW7 2AZ, United Kingdom. G.-Z. Yang is with the Institute of Medical Robotics, Shanghai Jiao Tong University, China.}}
\date{March 2020}

\begin{document}

\maketitle

\begin{abstract}
With increasing global age and disability assistive robots are becoming more necessary, and brain computer interfaces (BCI) are often proposed as a solution to understanding the intent of a disabled person that needs assistance. Most frameworks for electroencephalography (EEG)-based motor imagery (MI) BCI control rely on the direct control of the robot in Cartesian space. However, for 3-dimensional movement, this requires 6 motor imagery classes, which is a difficult distinction even for more experienced BCI users. In this paper, we present a simulated training and testing framework which reduces the number of motor imagery classes to 4 while still grasping objects in three-dimensional space. This is achieved through semi-autonomous eye-in-hand vision-based control of the robotic arm, while the user-controlled BCI achieves movement to the left and right, as well as movement toward and away from the object of interest. Additionally, the framework includes a method of training a BCI directly on the assistive robotic system, which should be more easily transferrable to a real-world assistive robot than using a standard training protocol such as Graz-BCI. Presented results do not consider real human EEG data, but are rather shown as a baseline for comparison with future human data and other improvements on the system.
\end{abstract}

\section{Introduction}
In recent years, Brain-Computer Interfaces (BCI) have been commonly proposed as a method of aiding disabled and healthy individuals with communication, brain state analysis, and device control \cite{Birbaumer2006BreakingControl}. Many efforts are focused on non-invasive methods of decoding brain signals based on electroencephalographic (EEG) recordings of motor imagery (MI) tasks, however real-time control of assistive robots through these methods remains challenging. In order to ensure accurate MI classification, each user needs to complete several intensive training sessions that match their particular EEG signals to a classifying algorithm \cite{Meng2016NoninvasiveTasks}.\par
There are several limitations of existing MI training protocols which make them difficult both for users and for translation to a robust assistive technology \cite{Jeunet2016WhyStudy}. Two of the main limitations are that training happens in an isolated, ideal environment for BCI while external environments are more dynamic and unpredictable, and that robotic control is often complex, with too many degrees of freedom to be directly handled by a MI-based BCI. 
Other limitations include that classifiers do not automatically update to incoming data, though there are several works which address this issue \cite{Freer2019AdaptiveProtocols}. \par
\begin{figure}
    \centering
    \includegraphics[width=\linewidth]{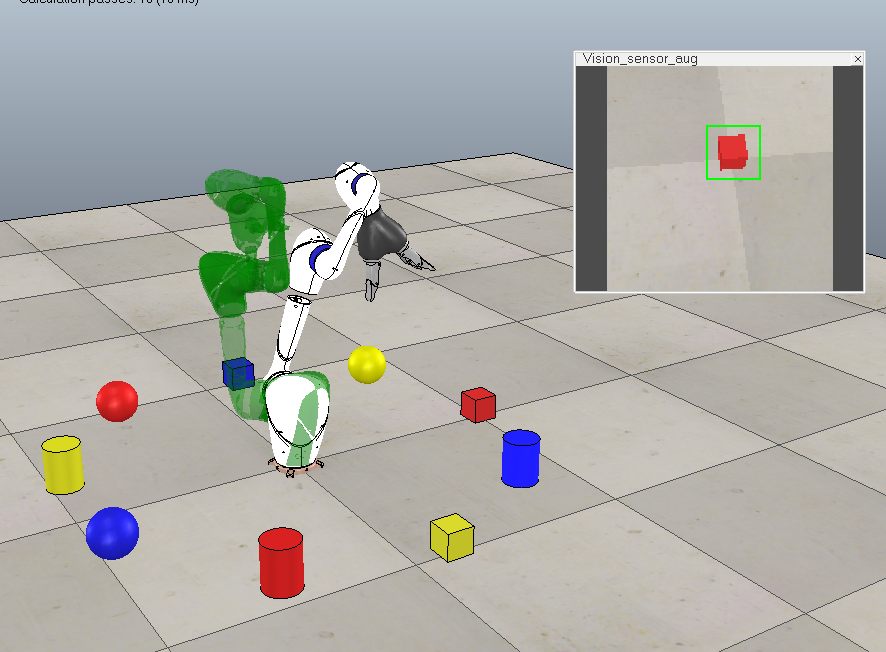}
    \caption{A picture of the MIndGrasp Trainer. A ``training" robot (green) moves while a user attempts to follow it by controlling the ``user" robot (white). EEG motor imagery data is recorded from the user, while augmented visual feedback is provided (top left). During testing, the ``training" robot is removed, and the user is asked to control the robot to grasp one of the 9 surrounding objects.}
    \label{fig:system_pic}
\end{figure}
During the beginning stages of training, using a plain background and minimalist design such as in the BCI-Graz protocol \cite{Pfurtscheller2001MotorCommunication} is suitable. This design ensures that external stimuli are largely reduced, so researchers can be fairly certain that the acquired EEG data is mostly related to the user's own thoughts and not other factors. However in daily life, visual, auditory and other stimuli are abundant, and have a significant effect on neural signals, which will in turn affect the control of the assistive robot. Therefore, in order to translate assistive robotic control into real environments, later stages of training should also be performed in environments with more stimulation that is similar to the environment in which the BCI will eventually be used. For this reason, it may be more useful to train a BCI in a simulated environment that closely resembles real-world assistive scenarios.\par
Another reason why it is difficult to translate MI control to real robotic systems is the gap between the robotic degrees of freedom (DoF) and how many DoF the user can reliably control at a time using an MI-based BCI. For example, most MI paradigms only consider up to 4 classes, but in order to directly move a robot in 3D space you need at least 6 classes corresponding to up, down, left, right, forward, and backward, respectively.  Even with 6 classes, the orientation of the robotic end-effector is not considered at all, and neither are other options such as when to grasp an object, grasp type, and so on. 
While there has been some research related to MI control for 3D movement, this either requires intensive training protocols \cite{McFarland2010ElectroencephalographicMovement} or autonomous movement in one DoF \cite{Lafleur2013QuadcopterInterface}, indicating that some semi-autonomous control may be necessary for more complex 3-dimensional operations. \par
In this paper, a new framework (MIndGrasp) for transferrable training of MI-based assistive robotic control is proposed which addresses the current limitations of MI training protocols. The platform includes a simulated V-REP training environment in which a ``training robot" moves to provide prompts to the user, and feedback is provided to the user by movement of a semi-autonomously controlled robot. The framework additionally includes a testing scenario for users to practice before moving into the real world. The framework is evaluated at a baseline level, and is largely presented to be built upon in the future.

\section{Related Work}
Recent research has shown that controlling a robotic arm using EEG motor imagery is possible \cite{Meng2016NoninvasiveTasks}, however these studies typically only move the robot in 1 to 2-dimensional space at a time, and the training protocol has been similar to the traditional Graz-BCI framework \cite{Pfurtscheller2001MotorCommunication}. In this framework, arrows appear on screen, pointing in one of four directions which indicate to the user how they should think in order to achieve the correct brain waves. Other training protocols have considered different types of visual prompts and feedback \cite{Hwang2009Neurofeedback-basedBCI}, but have still been mostly constrained to variations on the Graz-BCI framework \cite{Jeunet2016WhyStudy}. Beyond this, to the authors' knowledge there are no broadly used training protocols that can be applied to MI-based assistive robotic control. \par
For uninhibited and self-paced 3-dimensional MI control, research has been less common, though there have been several studies investigating this \cite{Meng2018Three-dimensionalTasks}. In an impressive paper, McFarland et. al. used 6 MI classes to provide 3-dimensional movement of a character in a virtual environment \cite{McFarland2010ElectroencephalographicMovement}. However, the training protocol for this task was intensive, requiring up to more than 50 training sessions amounting to over 20 hours of data for some subjects. While some research has attempted to reduce training time through adaptive protocols \cite{Freer2019AdaptiveProtocols} and data augmentation \cite{Freer_2020}, it is not clear how well these methods will perform as additional classes are added. \par
Another group has controlled a quadcopter in 3D space using MI with a reductionist training and processing algorithm \cite{Royer2011EEGSpace,Lafleur2013QuadcopterInterface}. However in this case, one of the dimensions was a constant forward-moving velocity while 4-class MI controlled the drone's yaw and altitude. This presents the idea of using semi-autonomy to achieve complex control with a simplified user interface, which is a concept also utilised in the control strategy of the proposed paper. \par
Lampe et. al. similarly used a semi-autonomous grasping procedure to test motor imagery control \cite{Lampe2014AGrasping}. Their BCI was trained using the Graz-BCI method, and the robotic grasping method seemed to be almost entirely autonomous, as learned through reinforcement learning, with the users mostly correcting erroneous movements \cite{Lampe2014AGrasping}. Ying et. al. primarily used computer vision and the grasp planner MoveIt! to iteratively provide different grasping options to the user, which were then selected via cursor movement on a screen by the user's operation of the BCI \cite{Ying2018GraspingSelection}. However, to date there are no MI control mechanisms for assistive robots which give users the ability to smoothly start, stop, select objects and correct throughout the grasping process.\par

\section{Framework}
Presented here is a framework to efficiently and robustly train and test an EEG MI classifier that can be more directly applied to assistive robotic control in real world environments.
The eventual vision for this framework is to use a single run of the MIndGrasp Trainer, automatically adapting the BCI classifier via the method described in \cite{Freer2019AdaptiveProtocols} until the user is able to achieve acceptable practical results on the simulator. However, because initially training in the dynamic environment may create too much noise in the EEG signal, a proposed graded training protocol can be seen in Fig. \ref{fig:MIndGrasp_training_framework}. In this protocol, data is initially collected from the standard Graz-BCI scenario, then feedback is presented to the user, and then users switch to the MIndGrasp Trainer, collecting training data directly related to MI-based simulated robotic control. The dataflow for each step of the process is additionally shown in Fig. \ref{fig:MIndGrasp_training_framework}.\par

The MIndGrasp simulator consists of five main parts, which will be discussed in the following subsections, and can be explored in more detail in the publicly available code\footnote{https://github.com/dfreer15/BCIRobotControl}:
\begin{enumerate}
    \item All objects in the simulator and their defined dynamic relationships. This includes the robot, all objects for the robot to grasp, and the training robot which provides a reference for the user during the training protocol;
    \item The control strategy of the robot, which manifested as a Finite State Machine, utilising both BCI and Computer Vision as inputs for semi-autonomous control;
    \item The BCI control algorithm, which converts users' brain signals into robotic movement. The BCI is the only direct input the user has to control the robot;
    \item The Computer Vision (CV) system, which aids in the semi-autonomous control of the assistive robot and allows augmented reality to be displayed to the user;
    \item The MIndGrasp Trainer, which utilises all of the previously mentioned components, and additionally includes a ``training" robot that provides the prompts for a user's MI-based control during training. 
\end{enumerate}

\subsection{Robotic Components and Physical Models}
V-REP (now CoppeliaSim) was used to create a simulator that could present our vision for this control strategy. Our framework was designed to work with multiple types of robots and grippers, some of which are automatically available in V-REP, and some of which were manually built in V-REP using acquired CAD models. The robots considered for use in this framework include the Hamlyn Active Arm (HAA) \cite{Wisanuvej2016}, Jaco, KUKA KR-10, and KUKA LWR. The grippers considered include the Jaco gripper, SVH, Barrett Hand, and the Salford Hand \cite{Mahboubi2018VariableHandling}. 

\subsection{Finite State Machine}
\label{subsubsec: FSM}
\begin{figure*}
  \begin{center}
  \includegraphics[width=\linewidth]{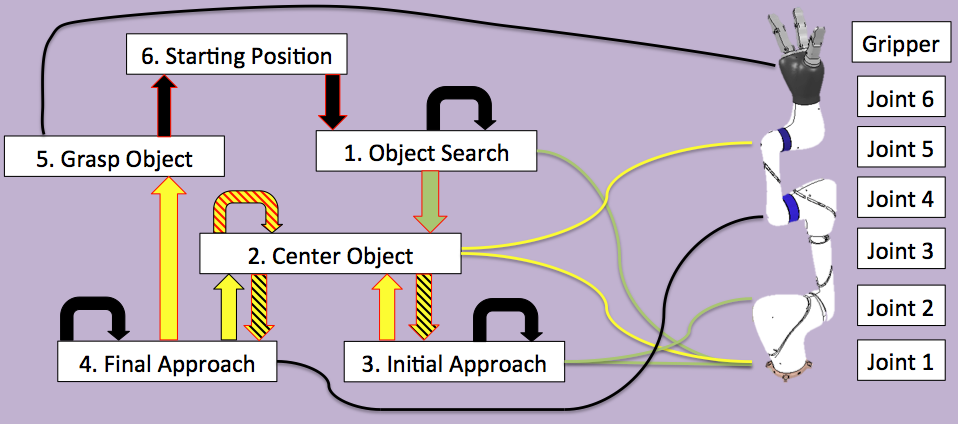} 
  \caption{The finite state machine used to control the approach and grasping of the robot in the simulator. The main pathway through the machine begins with Object Search, and progresses through the arrows lined in red. Each state change is determined by the BCI (green arrows), computer vision (yellow arrows), or their default pathway (black arrows). Pathways that may depend on multiple parameters are striped. Also pictured is the Hamlyn Active Arm, with lines explaining how each joint is affected when the robot is in a given state, similarly with the BCI (green), computer vision (yellow), or through default parameters (black).}
  \label{fig:FSMandRobot}
  \end{center}
\end{figure*}

In the studies presented here, the HAA was used along with the 3-fingered Jaco gripper. The HAA is a 6-DoF robot previously considered for use in robotic surgery \cite{Wisanuvej2016}, though it could be adapted for multiple applications. While the framework should largely still work for the variety of robots and grippers mentioned above, some parameters and the developed training protocol were most accurately implemented for these chosen robotic components.\par
Three different primitive object shapes were used for the majority of testing, which were cube, cylinder, and sphere. Three colours of each shape were used: red, yellow, and blue. These objects were chosen for computational and logistical simplicity, though future iterations of the framework should include more daily objects.
The physics engine used was the default for V-REP, which was based on Bullet 2.78.

Control of the selected robot and gripper was achieved through a Finite State Machine (FSM) with six states. In general, the first step was to search for the desired object by mentally turning the robot to the left or right. With the correct object in the robot's field of view, the user should imagine the opening and closing of both of their hands to begin the robot's approach. If any changes or errors occured during the approach, the user was able to correct these errors by moving to the left, right, or away from the desired object to recenter it in the field of view. The user could additionally move the robot more quickly toward the desired object if desired. When the robot's end-effector got close to the desired object, the autonomous grasping phase took over and the robot completed the grasp without any additional input from the user. A graphical representation of the robotic control system can be seen in Fig. \ref{fig:FSMandRobot}. The different states of the FSM are:
\begin{enumerate}
    \item Object Search: The robot does not move unless the BCI detects a command from the user. The robot will move left or right through the rotation of Joint 1 if left or right motor imagery is detected, and will enter into the ``Initial Approach" state if both hands motor imagery is detected, after fully centering the object. 
    \item Center Object: The robot's first and fifth joints move based on the CV system, ensuring that the current object of interest remains in the center of the robot's eye-in-hand field of view. After this, the robot will either return to the state it was previously in, or fully recenter the object before moving on to the next state.
    \item Initial Approach: Through movement of Joint 2, the long arm of the robot begins move downward, approaching the object of interest. Both hands and both feet motor imagery will move the long arm forward or backward, respectively. Left or right hand motor imagery turns joint 1 of the robot in the desired direction while stopping the approach, and may also change the object of interest. If no motor imagery class reaches above a certainty threshold, the default movement will be a slow approach toward the determined object of interest. When the object of interest is close to the end-effector, or before the robot's joint angles reach an unfavorable position, the final approach will begin. 
    \item Final Approach: The robot's long arm stays stationary while the short arm moves the end effector toward the object. This step is autonomous, with no BCI command from the user having any effect. The Grasp Object state is then reached depending on the CV system.
    \item Grasp Object: A signal is sent to the gripper which indicates that it should close. The final grasp is binary, either fully opening to its initial starting position, or closing with constant joint torques. Several seconds after closure, the robot enters State 6.
    \item Return to Starting Position: All joints turn toward their starting position until they reach within an acceptable threshold. The starting position is with all joints straight except for Joint 5, which angles the gripper, and therefore the eye-in-hand camera, toward the objects. When all joints have returned to the starting position, the ``Object Search" state begins again.
\end{enumerate}
In states 3 and 4, if the object of interest moves out of the center of the robot's field of view, the Center Object state (State 2) will be entered. In addition, when switching between all states except for State 2, the robotic system locks, stopping movement of all of its joints, before continuing on to the next state.

\subsection{Brain Computer Interface Control} \label{sec: methods_bci}

\begin{figure}
    \centering
    \includegraphics[width=\linewidth]{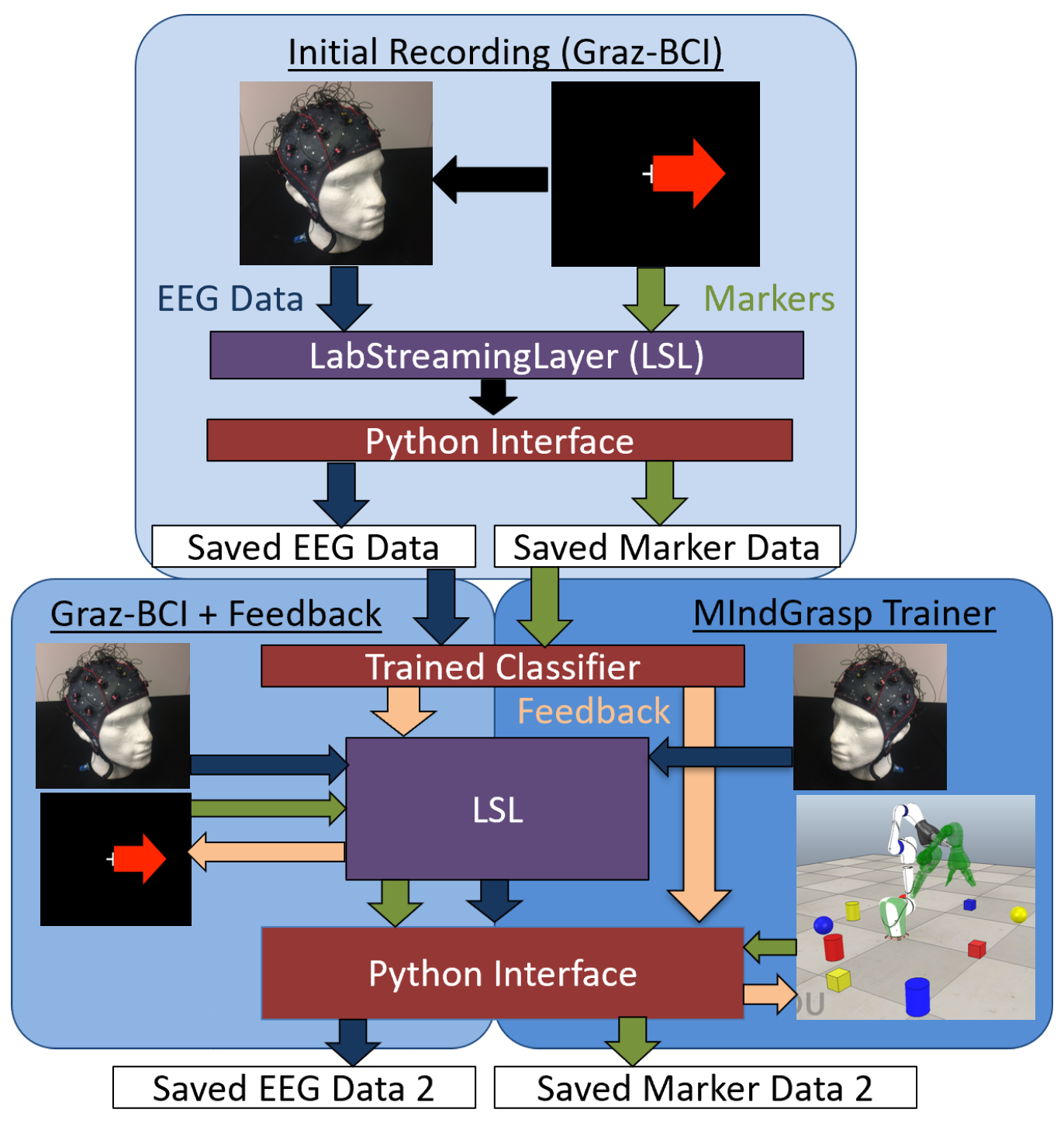}
    \caption{Dataflow for the proposed training framework in the MIndGrasp system. Initial data is recorded using the standard Graz-BCI protocol. After this, either feedback is added into the Graz-BCI protocol or the MIndGrasp Trainer is used.}
    \label{fig:MIndGrasp_training_framework}
\end{figure}

EEG data was collected from a 32-channel g.tec g.Nautilus wet cap and was streamed to the overarching program using LabStreamingLayer (LSL). After passing through the classifier, high frame rate predictions and certainties were sent to the V-REP simulator via its python interface, in contrast to the Graz-BCI + feedback scenario, which used the LSL interface for both steps. The dataflow for all three stages of training can be seen in Fig. \ref{fig:MIndGrasp_training_framework}. \par 
The MI predictions were only utilised when the robotic simulator was in States 1, 2, and 3, as mentioned in Section \ref{subsubsec: FSM}. The 4 classes, consisting of left (0), right (1), and both (2) hands motor imagery, and both feet (3) motor imagery, generally corresponded to robotic movement to the left, right, toward an object, or away from an object, respectively. Classification was performed using a Riemannian MDM classifier, which, among other Riemannian classifiers \cite{Yger2017RiemannianReview}, has shown significant promise in recent years \cite{Freer_2020, Barachant2012MulticlassGeometry}.\par
The BCI controlled the simulated robot at the task level, essentially being used to find and select desired the object, then to tell the robot to grasp while ensuring that the desired object remains the robot's object of interest. The velocity of the robot joints were set based on the certainty that the classifier chose the correct MI class. Because the classifier chose the class with the smallest distance to the Riemannian mean, the certainty value (cert) was calculated by comparing how much smaller this distance was than the distance to each of the other classes. The method used to determine this certainty value can be seen in Equation \ref{eq:cert_mindgrasp}. 
\begin{equation}cert = \dfrac{\sum_{i=1}^C\delta_{Ri}}{C} - min(\delta_{R1...C}) \label{eq:cert_mindgrasp}\end{equation}
The calculated certainty value was scaled and set to the joints indicated by the green lines as indicated in Fig. \ref{fig:FSMandRobot}. More specifically, the left and right classes turned Joint 1 while searching for an object (State 1) and during the initial approach (State 3), while the both hands and both feet classes affected Joint 2 only when the robot was in State 3. If both feet was the determined class, then Joint 2 would begin to move backward, while if both hands was the determined class, Joint 2 would move the end-effector more quickly toward the current object of interest. In contrast, in State 1, both hands motor imagery was only used to progress to the next state, which would be State 2 (centering the object in the camera's view), then immediately State 3 (approaching the object). Additionally, in State 2 (centering the object of interest) the determined MI class informed the CV system which object might be the object of interest, which will be discussed in the following subsection.

\subsection{Computer Vision System}
The CV system for this framework made use of the bright color of each object in order to determine the object's location. This was achieved by simply thresholding the RGB values of each pixel
and marking the position of each pixel that matched one of the known colors from our set. The center of a given object was determined to be the midpoint between the maximum and minimum object pixel position in the two-dimensional image coordinate frame. The number of pixels that matched a given color was also recorded in order to determine the distance between the end-effector's eye-in-hand camera and the object of interest. This was only possible because the size of the objects were known beforehand, so with novel objects a depth camera would likely be necessary during the initial approach.\par 
Within States 3 and 4, the FSM automatically temporarily moved to State 2 to center the view of the robot on the object of interest using direct proportional control of motors 1 and 5 (Fig. \ref{fig:FSMandRobot}). Additionally, when switching between states the vision system was used to fully center the object in the robot's eye-in-hand view, stopping only when the object's pixel area center was within a threshold of the center of the image view.\par
If more than one object was in view of the camera, the CV system also ranked the likelihood that a given object was the true ``object of interest" based on how close the object was to the center of the image, and the total number of pixels that object contains. This determination was very important, because the choice of the ``object of interest" determined which object the robot centered itself on in state 2 above. In addition, if either the ``left" or ``right" class was determined by the BCI, the point considered as the ``center" of the image was shifted to 1/4 or 3/4 of the way across the image, from left to right, respectively, rather than 1/2 of the way if neither of these classes was determined. The chosen object of interest was also indicated to the user via augmented reality on the display of the eye-in-hand camera. The augmented reality chosen was a green box around the object of interest, generated by OpenCV.\par

The CV system was additionally used just before state 4, in which a depth image of the object of interest was taken, and the shape of that object was determined via a 5-layer Deep Convolutional Neural Network (DCNN). In the future, this shape determination step could be used to determine grasp type for the end-effector \cite{Wang2019Vision-basedLiving} or determine the final autonomous grasping strategy. The DCNN was trained on about 500 depth images collected in the simulator with the robot in this intermediate state, consistently achieving over 90\% accuracy on the validation set. The DCNN was implemented in Tensorflow using Adam optimizer, with sparse categorical crossentropy loss. The class determined by this network changed the parameters used during state 3 of the system, and also determined the shape of the object which was grasped for results comparison. \par
While the visual control system presented here needs improvements to robustly work in a real-world environment, it was adequate to test our current framework for BCI control in simulation. However, the entire system is largely modular, so it should be trivial to upgrade it to include concepts such as classification and localization using deep learning \cite{Wang2019Vision-basedLiving}, though speed may be an important factor with more complex architectures. Additionally, as the robot autonomously completes the grasp from step 4 onward, other more complex grasping techniques could be implemented that have been learned through techniques such as deep reinforcement learning \cite{Kalashnikov2018QT-Opt:Manipulation, Zeng2018LearningLearning}, rather than just tuning a few parameters. However, each of these implementations will also need to consider the possibility for real-time human-centered control. The focus of this work was primarily to bring motor imagery training and control into more realistic daily environments, so these methods and other state-of-the-art CV systems were not fully investigated. \par

\subsection{MIndGrasp Trainer}
In addition to providing a way to practice and test MI-based robotic control, the MIndGrasp simulator includes its own method for training an MI classifier. Similar to Graz-BCI, the MIndGrasp Trainer provides timed prompts to a user, indicating how they should change their mental strategy. However, in the MIndGrasp Trainer the prompts are not arrows pointing in a given direction, but are instead manifested in the movement of a ``training" robot that behaves similarly to the user-controlled robot. In this way, the user trains themselves and the classifier by attempting to follow the movement of the training robot as closely as possible. \par
The MIndGrasp Trainer makes use of exactly the same environment as has been described thus far, but also adds a training robot into the V-REP scene as a green semi-transparent replica of the HAA, as can be seen in Fig. \ref{fig:system_pic}. It is controlled autonomously through Lua code written in the V-REP scene. This code randomly selects one of the four classes, moves the robot that direction for 2 seconds, then stops moving and waits for the user for 2 seconds before choosing another class and continuing to prompt the user through its movement. This timing was chosen to be approximately the same as the Graz-BCI training protocol. 
Prompts are sent to the python interface with a timestamp and saved in the same way as in the Graz-BCI training protocol which was previously implemented for data collection \cite{Freer2019AdaptiveProtocols}. As a result, these prompts can be read directly into the next training or testing phase to provide labels for the training data.
The user attempts to follow the training robot with the real robot using the same control methodology described in the above subsections. In contrast, the training robot is simply controlled through joint velocities. This difference in control will produce an offset between the two robots, though higher MI accuracy should still lead to decreased positional error of the end-effector. 
If the end-effector of the training robot reaches beneath a set height of 0.1 meter, then both robots would be sent a signal to return to their initial position. Similarly, if the human-controlled robot enters into its final grasp phase, then both robots are also sent a signal to return to their initial position.

One benefit of this training strategy is that it should be easy for assistive robot users and researchers to immediately evaluate the task-based ability of the control method based on how closely users are able to follow the training robot. This training method can also allow for quantitative practical accuracy metrics such as total distance between the end-effectors of the two robots for a given trial, which could be able to more accurately predict task completion than directly using MI classification accuracy.

\section{Experimental Methods}

\subsection{Set Locations}
To ensure that the BCI could, in fact, control the robot to grasp different objects in the scene with the previously described strategy, 9 graspable objects were placed around the robot, 0.5 m away from its base. The 9 objects were red, blue, and yellow versions of the three primitive shapes of cube, cylinder, and sphere, respectively. This allowed for fairly easy CV classification using RGB and depth information. The objects' placement can be seen in Fig. \ref{fig:mindgrasp_results}.\par
For this process, a classifier was first trained on EEG data from an unworn cap using the Graz-BCI scenario and then the MIndGrasp Trainer. This trained classifier was used with real-time EEG data to generate classes and certainties to control the robot in the testing scenario. A CV classifier was then trained on depth images from just before the final grasp stage of previous trials in order to make the final object classification before completing the grasp. \par
From here, a single grasp trial would begin, in which the ``desired" object was randomly selected and displayed in the bottom right hand corner of the simulator, and then the robot would enter the Finite State Machine and follow the protocol described in \ref{subsubsec: FSM}. Once the robot finally attempted a grasp on one of the objects in the scene, data about that grasp was recorded, including the ``desired" object, the object that was grasped (as determined by the trained depth classifier and the final RGB image of the object before the final grasp), and the time taken to complete the grasp. From here, the robot and all objects were reset to their initial position and a new grasp trial began. 160 grasp attempts were considered for this experiment.\par

\subsection{Random Locations}
To test whether the grasping control method could still work in non-ideal conditions, three primitive shapes were placed in random locations around the HAA in the V-REP environment. The objects were placed within a suitable radial range of the robot, as objects placed outside of the reachable workspace could not feasibly be grasped. The distances chosen for this initial testing were between 0.2 and 0.7 m away from the base of the robot. Random objects were attempted to be grasped 507 times when placed in random locations. 
In this scenario, no BCI control was used, and the robot searched for objects by turning to the left until it found one in its field of view, then autonomously completed the grasp using only computer vision. Only one randomly selected object was visible during a given trial, which ensured that the robot could only find and attempt to grasp this object.

\section{Results}

\subsection{Set Locations}
Out of 160 grasping trials with the objects in their set testing locations, 57 grasps were successful, for a rate of 35.6\%.
\begin{figure}
    \centering
    \includegraphics[width=\linewidth]{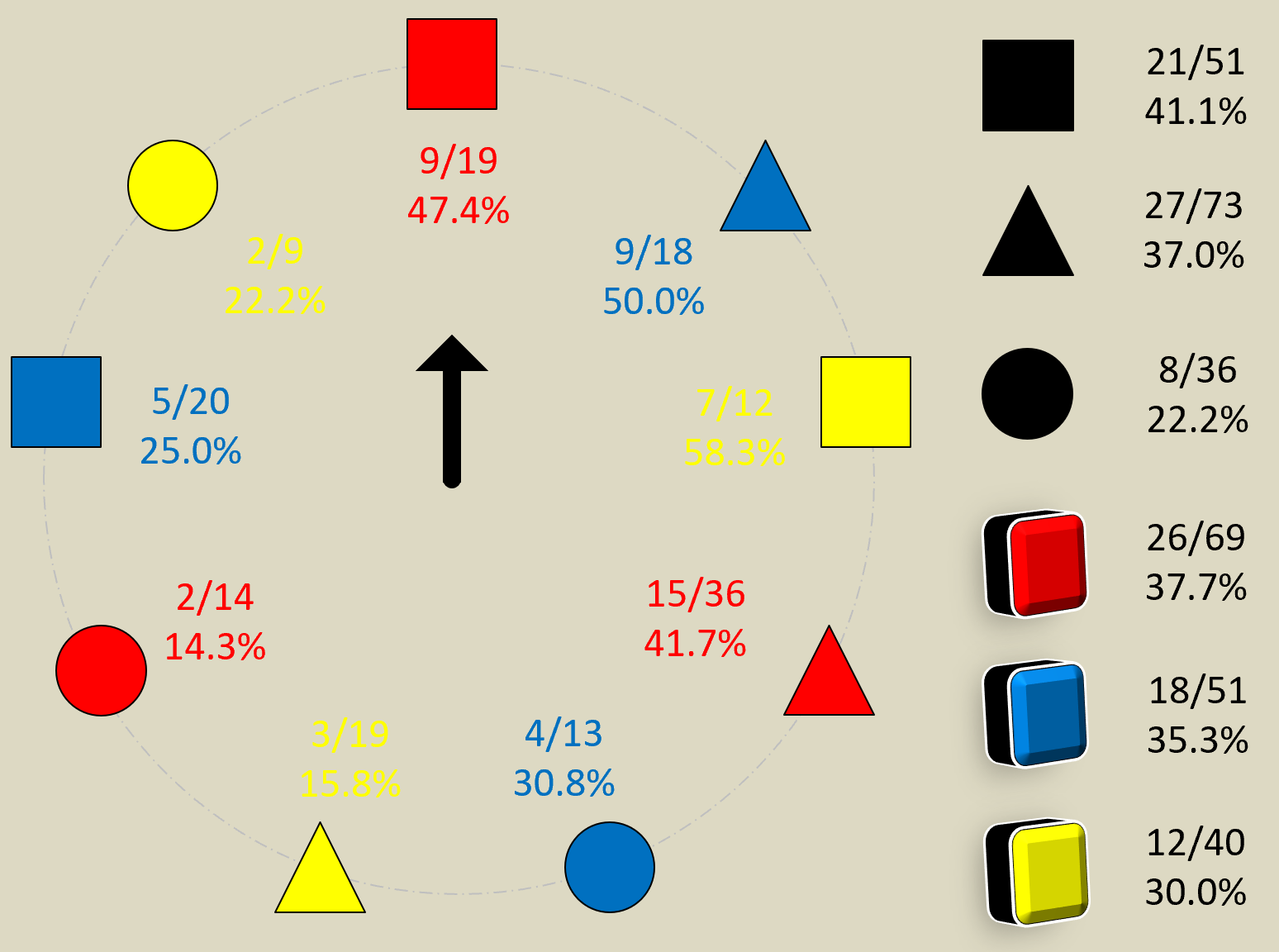}
    \caption{Grasping results of the MIndGrasp simulator with the objects in their set locations. The squares, triangles, and circles represent the set location of a cube, cylinder, or sphere, respectively, while the colors represent the object color. The black arrow shows the initial direction of the robotic arm.}
    \label{fig:mindgrasp_results}
\end{figure}
The types of attempted grasps were fairly well distributed between the different object types and locations. With 160 grasp trials, the expected statistical result of random grasping would have been about 18 attempted grasps for each object type. Most objects were close to this, as can be seen in Fig. \ref{fig:mindgrasp_results}. The most commonly grasped object was the red cylinder, with 36 attempted grasps and 15 successful grasps (41.7\%), while the least commonly grasped object was the yellow sphere, with only 9 attempted grasps; 2 were successful.\par
The spheres were grasped the least often, with only 36 attempted grasps and 8 successful grasps. The reason for this was likely because although the spheres were placed in their initial position at the beginning of each trial, the nature of the physics engine caused the spheres to have some random movement, shifting their location away from their initial position. This was solved through the addition of invisible "bins" that kept the spheres close to their starting position, but the spheres still had a random velocity throughout the trials. This could lead to a reduction in attempted grasps because the robot may not be able to visually follow a moving object as well as a static one, and could also result in a lower percentage of successful grasps because the object may move outside of the optimal position for the gripper. \par

Additionally, the random EEG data selected the correct object 23 of the 160 trials, for a rate of 14.4\%. This is slightly better than the expected result of 11.1\%, but should still be much lower than if a human is truly able to control the robot. Of these correct selections, only 5 were successful grasps, leading to only 3.1\% of trials for which the correct object was chosen and successfully grasped. Once human data is collected it can be compared to these values for verification of successful control.

The average time taken to complete a single grasp in this scenario was 246.4 seconds. This is a long time necessary just to simply grasp an object, but the main reasons for this should be solved with real human data as opposed to streamed ambient data of an unworn cap. For example, if the classifier had been trained on data with four distinct classes, the classifier's ability to discriminate between these classes would have been better, leading to more consistency between neighboring time windows, and a higher certainty value. With this, the robot wouldn't be changing direction as much (such as moving forward, then backward, or left, then right), and would also move with a higher velocity. Therefore, the time taken to complete a single grasp trial would be reduced. \par

\begin{figure*}
    \centering
    \includegraphics[width=0.325\linewidth]{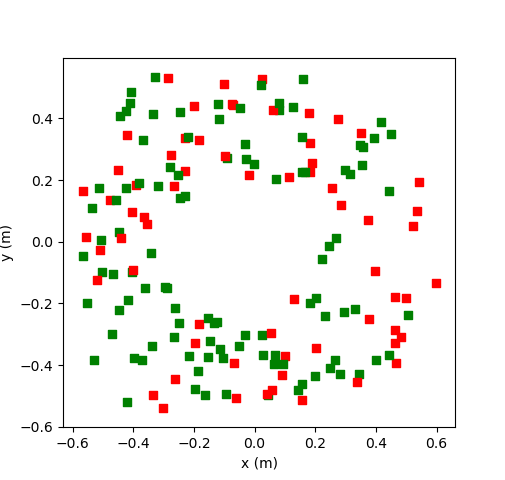}
    \includegraphics[width=0.328\linewidth]{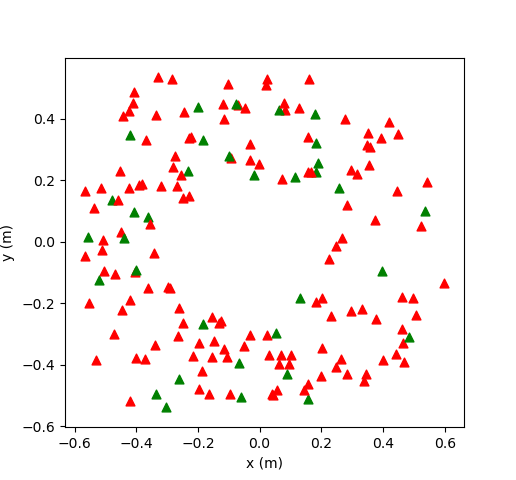}
    \includegraphics[width=0.325\linewidth]{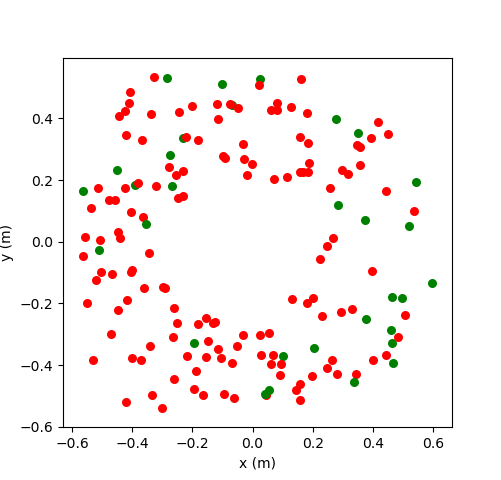}
    \caption{Scatter plots displaying successful (green) and failed (red) grasps depending on location for cube (left), cylinder (middle), and sphere (right).}
    \label{fig:rand_grasp_success}
\end{figure*}

Though the presented data was not collected on real subjects, testing with an unworn cap verifies that the flow of data works, and that a variety of results can be produced due to changing real-time EEG data. The results of these experiments can also serve as a baseline for comparison with real human EEG data and with more sophisticated control strategies that could be included in future iterations of the MIndGrasp Framework. \par

\subsection{Random Locations}
With 507 grasp trials with the objects placed in random locations, the previously described method of grasping had a success rate on all objects of about 33.5 percent. The highest success rate was seen when attempting to grasp the cube, with an overall success rate of about 60.5 percent, while the corresponding rates for cylinder and sphere were 17.6 and 13.3 percent, respectively. Fig. \ref{fig:rand_grasp_success} shows the initial placement of each successful and unsuccessful grasp for cube, cylinder, and sphere, respectively. \par 
These rates are unacceptable for daily assistance, so the final autonomous phase of grasping should be significantly updated. This grasp policy could be governed by deep reinforcement learning, with considerations into object type using the depth camera, and could additionally use the gripper joint torque values or tactile sensing to ensure the stability of a grasp before lifting the end-effector.
\section{Discussion}

While a significant amount of work toward this framework was completed, there is still much room for further improvement. For example, the transfer between different training phases has not been tested thoroughly in terms of the real-time application of human EEG signals. For example, preliminary tests showed that achieving high classification accuracy in the Graz-BCI scenario did not directly translate to successful robotic control in the MIndGrasp simulator. One way of easing the transfer between Graz-BCI training and the MIndGrasp Trainer may be to randomise the color or shape of the stimulants and feedback within Graz-BCI, which could provide a more robust mapping of EEG data to MI control signals. This could serve as a sort of data augmentation with its foundations in the stimulants rather than in the data itself. However, more experiments must be conducted to explore this further.

The performance of the MIndGrasp Simulator must also be improved. One of the benefits of using a task-based training method may be the positive reinforcement that somebody could receive from successfully completing a grasping task. If the user performs the task adequately, but the robot is unsuccessful in its grasp, this positive reinforcement may be lost. One method of solving the problem would be to programmatically couple the grasped object to the end-effector once the final stage of the grasp is executed, but this would ignore the true physical properties of the robot. Other ways of improving grasping performance would require a new control policy for the last autonomous stage of the grasp. This control policy could be determined through vision-based grasp planning, reinforcement learning, or through intelligent shaping and control of the end-effector \cite{Wang2019Vision-basedLiving}.


\section{Conclusion}

In conclusion, this paper proposed and characterised a framework for EEG MI-based robotic training and control which is more transferrable to daily assistive robotic control. In addition, a new MI-based control methodology is proposed which can translate 4 MI classes into semi-autonomous object grasping in 3D space. 
Preliminary testing of the MIndGrasp system indicates that the robot is responsive to classified EEG data, and is able to semi-autonomously achieve grasps on objects within its range. However, the grasping accuracy is still not as high as desired. While not all aspects of the framework have been verified, the platform will be made open-source for future research to further characterise and build upon. The hope is that this work will provide future BCI-controlled assistive robots with a basic framework to train and test with new users before moving on to real-world environments. Such an intermediary step is necessary to verify EEG-based MI control of assistive robots in dynamic environments. \par

\bibliographystyle{IEEEtran}
\bibliography{MIndGrasp}

\end{document}